\documentclass[preprint2]{aastex}
\usepackage{color}
\usepackage{float}
\usepackage{graphicx}
\usepackage{wrapfig}

\slugcomment{Not to appear in Nonlearned J., 45.}

\shorttitle{A new sdO+dM eclipsing binary}
\shortauthors{Derekas et al.}

\begin{document}

\title{A new sdO+dM binary with extreme eclipses and reflection effect}

\author{A. Derekas\altaffilmark{1,2}, P. N\'emeth\altaffilmark{3}, J.
Southworth\altaffilmark{4},  T. Borkovits\altaffilmark{5,1}, K.
S\'arneczky\altaffilmark{2}, A. P\'al\altaffilmark{2}, B. Cs\'ak\altaffilmark{1}
, D. Garcia-Alvarez\altaffilmark{6,7,8}, P. F. L. Maxted\altaffilmark{4}, L. L.
Kiss\altaffilmark{2,9}, K. Vida\altaffilmark{2},  Gy. M.
Szab\'o\altaffilmark{1,2},  L. Kriskovics\altaffilmark{2}}
\altaffiltext{1}{ELTE Gothard Astrophysical Observatory, H-9704 Szombathely,
Szent Imre herceg \'ut 112, Hungary; E-mail: derekas@gothard.hu}
\altaffiltext{2}{Konkoly Observatory, Research Centre for Astronomy and Earth
Sciences, Hungarian Academy of Sciences, H-1121}
\altaffiltext{3}{Dr. Karl Remeis-Observatory \& ECAP, Astronomisches Inst., FAU
Erlangen-Nuremberg, 96049 Bamberg, Germany}
\altaffiltext{4}{Astrophysics Group, Keele University, Newcastle-under-Lyme, ST5
5BG, UK}
\altaffiltext{5}{Baja Astronomical Observatory of Szeged University, H-6500 Baja, Szegedi \'ut, Kt.
766, Hungary}
\altaffiltext{6}{Instituto de Astrof\'{\i}sica de Canarias, E-38205 La Laguna,
Tenerife, Spain}
\altaffiltext{7}{Dpto. de Astrof\'isica, Universidad de La Laguna, 38206 La
Laguna, Tenerife, Spain}
\altaffiltext{8}{Grantecan CALP, 38712, Bre\~{n}a Baja, La Palma, Spain}
\altaffiltext{9}{Sydney Institute for Astronomy, School of Physics, University
of Sydney, Australia}

\begin{abstract}
We report the discovery of a new totally-eclipsing binary ({\sc RA}=$\rm{06^{h}40^{m}29^{s}11}$; {\sc
 Dec}=$\rm{+38^{\circ}56\arcmin52\arcsec2}$; J=2000.0; R$_{\rm
{max}}$=17.2 mag) with an sdO primary
and a strongly irradiated red dwarf companion. It has an orbital period of
$P_{\rm {orb}}$=0.187284394(11)~d and an optical eclipse depth in excess of 5
magnitudes. We obtained two low-resolution classification spectra with
GTC/OSIRIS and ten medium-resolution spectra with WHT/ISIS to constrain the
properties of the binary members. The spectra are dominated by H Balmer and
He\,{\sc ii} absorption lines from the sdO star, and phase-dependent emission
lines from the  irradiated companion. A combined spectroscopic and light curve
analysis implies a hot subdwarf temperature of $T_{\rm eff}({\rm
spec})=55\,000\pm3000$\,K, surface gravity of $\log{g} ({\rm phot})=6.2\pm0.04$
(cgs) and a He abundance of $\log(n{\rm He}/n{\rm H})=-2.24\pm0.40$. The hot sdO
star irradiates the red-dwarf companion, heating its substellar point to about
$22\,500$\,K. Surface parameters for the companion are difficult to constrain
from the currently available data: the most remarkable features are the strong H
Balmer and C\,{\sc ii-iii} lines in emission. Radial velocity estimates are
consistent with the sdO+dM classification. The photometric data do not show any
indication of sdO pulsations with amplitudes greater than 7\,mmag, and
H$\alpha$-filter images do not provide evidence of the presence of a planetary
nebula associated with the sdO star.
\end{abstract}


\section{Introduction}{\label{Sec:intro}}

Hot subdwarf stars are located between the upper main sequence and the white
dwarf (WD) sequence in the Hertzsprung-Russell diagram. They are evolved,
core-helium burning, low-mass stars ($M\approx 0.5 M_{\odot}$) with very thin
hydrogen envelopes \citep{heb09}. Among hot subdwarfs, sdO stars ($T_{\rm eff} >
38000~K$) represent a significantly smaller fraction than sdBs ($T_{\rm eff} <
35000~K$). Spectroscopically, sdO stars show a large variety: the two main
groups are the H-rich (sdO) and He-rich (He-sdO) stars. \citet{str07}
showed that a strong correlation exists among surface temperature, He and C, N
abundances in He-sdO stars.

Canonical stellar evolution theory predicts that sdO stars evolve from sdB
stars. Whilst binarity is quite frequent among sdBs, with a binary fraction of
about 50\% \citep{max01,nap04}, the fraction of binary He-sdOs is very low
\citep{kaw15}. Many sdO binaries are associated with a planetary nebula (PN),
like UU Sge, V477 Lyr and BE UMa \citep{pol94a,pol94b,afs08}. The sdO stars in
these binaries are hotter and more massive stars in the immediate post giant
branch stage. Binaries with compact sdO stars evolve from sdB binaries and the
$\sim$120 Myr sdB lifetime is long enough for their PNe to attenuate and become
hardly detectable \citep{all15}. 

\citet{han02,han03} performed binary population syntheses and identified several
evolutionary channels that lead to the formation of hot subdwarf stars. In close
binaries that evolve through one or more common envelope (CE) phases \citep{pac76}, the secondary is engulfed by the atmosphere of the primary
while it is on the red giant branch. 
 
As the secondary spirals inward due to tidal friction, the red giant loses mass. 
By the end of the CE phase the primary loses most of its envelope and the binary 
orbital period shrinks to a few hours. If the core gained enough mass  for He 
ignition during the preceding evolution the primary experiences a He-flash and 
settles on the extreme horizontal branch. In case the core mass is insufficient for He burning, the 
primary evolves as a low-mass pre-white dwarf. 
In both cases the common envelope is ejected during the final stage of CE evolution 
and a very close binary remains \citep{taam06}.
While
this theory can explain the mass loss required for the formation of hot
subdwarfs, it also needs a precise timing between mass loss and the core helium
flash. Eclipsing hot subdwarf binaries with irradiated companions can give
insight into the details of these processes, making such binaries fundamental to
understand CE evolution.

Here we report the discovery of a new eclipsing binary with an sdO primary and a
strongly irradiated red dwarf companion. After describing the observations, we
discuss the spectral modelling and determination of the stellar and orbital
parameters from the spectroscopic and light curve analysis.

\section{Observations}{\label{Sec:obs}}

Konkoly J064029.1+385652.2 ({\sc RA}=$\rm{06^{h}40^{m}29^{s}11}$; {\sc
Dec}=$\rm{+38^{\circ}56\arcmin52\arcsec2}$; J=2000.0; R$_{\rm
{max}}$=17.2 mag; hereafter J0640+3856) was discovered serendipitously during
regular astrometric observations of minor planets. On one of the images the
object completely disappeared, suggesting a sudden deep eclipse. We started to
monitor J0640+3856 using the 0.6/0.9/1.8~m Schmidt telescope at Piszk\'estet\H o
Observatory. We took CCD photometric observations on seven nights between
December 2013 and February 2014 using Johnson/Bessell $V$ and Cousins $R_{\rm
C}$ filters, and also without filters. The telescope was equipped with an Apogee
ALTA-U 4k$\times$4k CCD camera. The observations revealed that the period is
0.187~d and there is a strong reflection effect with an amplitude of
$\sim$0.5\,mag. The sudden and deep primary minimum suggested that the system
might contain a hot and small primary star eclipsed by a cool secondary object.

We obtained further observations via the service program on the 4.2\,m William
Herschel Telescope (WHT) on La Palma. In February and March 2014 we observed two
primary eclipses with Sloan $r'$ and $i'$ filters and a secondary eclipse in
$i'$ band, using the ACAM imager \citep{ben08}. The data were reduced with standard
procedures. The flat minima indicated that the eclipse depths are 6 and 5 mag,
respectively, and the totality of the primary eclipse lasts for about 5.7\,min,
while the eclipse duration is approximately 24\,min.

On March 1, 2014, we obtained two spectra using the 10.4\,m Gran Telescopio Canarias (GTC) and OSIRIS
spectrograph\footnote{http://www.gtc.iac.es/instruments/osiris} on La Palma. Due to the highly variable seeing we used a 1.23 arcsec slit providing a dispersion of $\Delta\lambda\approx1.6$\,\AA/pixel and
$\Delta\lambda\approx4.2$\,\AA/pixel with the R2500V and the R1000B grisms,
respectively. These low- and medium-resolution spectra were taken between 3850-7400 \AA\ and 4400-6000 \AA, and reached a signal-to-noise ratio (SNR)\,$\sim$\,30 with 540\,s exposure times.

We acquired phase-resolved spectroscopy on March 6, 2014, using the WHT and the
dual-beam {\sc ISIS} spectrograph\footnote{http://www.ing.iac.es/Astronomy/instruments/isis}. This was operated with the R600B/R600R gratings and a 1 arcsec slit, providing a resolution of $\Delta\lambda=1$\,\AA/pixel in the blue and covering the 3800-5200\,\AA\ and 6200-7800\,\AA\ regions. Since the $H\alpha$ line was weak, we did not use the red region in the analysis. We took ten spectra between orbital phases $\varphi=0.47$ and $0.82$ with 600\,s exposure times. The average SNR of these spectra is 25. 

All spectroscopic data were reduced with our IRAF based data reduction pipeline.
Bias and flat field corrections were done with the {\sc ccdproc} task, raw
spectra and arc calibrations were extracted with {\sc apall} and wavelength
calibrated with the {\sc identify/reidentify} tasks. We identified $\sim$150
CuArNe lines in the WHT calibration data, that allowed for an eighth order
Legendre polynomial dispersion function with a root-mean-square (RMS) $\sim$0.1\,\AA.
The GTC HgArNeXe calibrations allowed us to identify $\sim$50-100 lines and use
a sixth order dispersion function with RMS$\sim$0.6\,\AA.

In April 2014 we obtained 1.5\,h CCD photometry with the 1\,m RCC telescope at
the Piszk\'estet\H o Observatory, using an FLI camera with a field of view of
9.4$^\prime$$\times$9.4$^\prime$, with a Sloan $r'$ filter and 2$\times$2
binning giving a  plate scale of 0.27$^{\prime\prime}$/pixel. In addition, we
obtained $\sim$21\,h of fast photometry (5\,s exposure times) in December 2014
and February 2015 using the same telescope with the OCELOT camera (an Andor
iXon+888 EMCDD camera) without filter, in order to search for any oscillations
from the sdO star. We did not detect any significant periodic signal with
amplitudes greater than 7\,mmag.

In order to detect whether any PN could be associated with the sdO, we took 27
H$\alpha$-filter images with 120\,s exposure times with the same telescope in
April 2014. The composite image did not show any sign of a PN.

\section{The orbital period}{\label{Sec:orbperiod}}

\begin{table*}
\begin{center}
\centering
\caption[]{\centering Times of minima of J0640+3856. }
 \label{Tab:ToM}
\begin{tabular}{lrll | lrll}
\hline
\multicolumn{4}{c|}{Primary minima} & \multicolumn{4}{c}{Secondary minima}\\
\hline
BJD (TDB) & Cycle  & std. dev. & instrument & BJD (TDB) & Cycle  & std. dev. & instrument \\
$-2\,400\,000$ & no. &   \multicolumn{1}{c}{$(d)$} & & $-2\,400\,000$ & no. &   \multicolumn{1}{c}{$(d)$} & \\
\hline
56684.301289 & -135.0 & 0.000004 & Schmidt R$_C$ & 56654.616610 & -293.5 & 0.000019 & Schmidt unfilt.\\
56684.488447 & -134.0 & 0.000004 & Schmidt R$_C$ & 56657.424647 & -278.5 & 0.000019 & Schmidt unfilt.\\
56691.417455$^*$&-97.0 & 0.000031& Schmidt V     & 56684.394740 & -134.5 & 0.000019 & Schmidt R$_C$\\  
56692.354351 &  -92.0 & 0.000004 & Schmidt V     & 56689.265080 & -108.5 & 0.000032 & Schmidt V\\
56693.290882 &  -87.0 & 0.000004 & Schmidt R$_C$ & 56691.325691 &  -97.5 & 0.000007 & Schmidt V\\
56695.350946 &  -76.0 & 0.000001 & WHT r'        &  56691.509921 &  -96.5 & 0.000017 & Schmidt V\\
56709.584600 &    0.0 & 0.000002 & WHT i'        & 56692.260600 &  -92.5 & 0.000019 & Schmidt V\\
56775.321665 &  351.0 & 0.000021 & RCC r'        & 56692.448527 &  -91.5 & 0.000024 & Schmidt V\\
57022.537442 & 1671.0 & 0.000005 & RCC unfilt.   & 56693.385087 &  -86.5 & 0.000023 & Schmidt R$_C$\\
57067.298585 & 1910.0 & 0.000002 & RCC unfilt.   & 56729.530347 &  106.5 & 0.000016 & WHT i'\\
57067.485826 & 1911.0 & 0.000003 & RCC unfilt.   & 57022.631307 & 1671.5 & 0.000021 & RCC unfilt.\\
57068.234937 & 1915.0 & 0.000003 & RCC unfilt.   & 57067.391972 & 1910.5 & 0.000015 & RCC unfilt.\\
57068.422263 & 1916.0 & 0.000002 & RCC unfilt.   & 57068.328437 & 1915.5 & 0.000013 & RCC unfilt.\\
             &        &          &               & 57068.515991 & 1916.5 & 0.000018 & RCC unfilt.\\
%\hline
%\multicolumn{4}{c}{Secondary minima}\\
%\hline
%56654.616610 & -293.5 & 0.000019 & Schmidt unfilt.\\
%56657.424647 & -278.5 & 0.000019 & Schmidt unfilt.\\
%56684.394740 & -134.5 & 0.000019 & Schmidt R$_C$\\
%56689.265080 & -108.5 & 0.000032 & Schmidt V\\
%56691.325691 &  -97.5 & 0.000007 & Schmidt V\\
%56691.509921 &  -96.5 & 0.000017 & Schmidt V\\
%56692.260600 &  -92.5 & 0.000019 & Schmidt V\\
%56692.448527 &  -91.5 & 0.000024 & Schmidt V\\
%56693.385087 &  -86.5 & 0.000023 & Schmidt R$_C$\\
%56729.530347 &  106.5 & 0.000016 & WHT i'\\
%57022.631307 & 1671.5 & 0.000021 & RCC unfilt.\\
%57067.391972 & 1910.5 & 0.000015 & RCC unfilt.\\
%57068.328437 & 1915.5 & 0.000013 & RCC unfilt.\\
%57068.515991 & 1916.5 & 0.000018 & RCC unfilt.\\
\hline
\end{tabular}
\end{center}
\begin{center}
\vspace{-6pt}
{\small {\bf Note.} Secondary minima, and the one primary minima denoted by $*$ were not
included in the period study.}
\end{center}
\end{table*}

\begin{figure*}
\includegraphics[width=\textwidth]{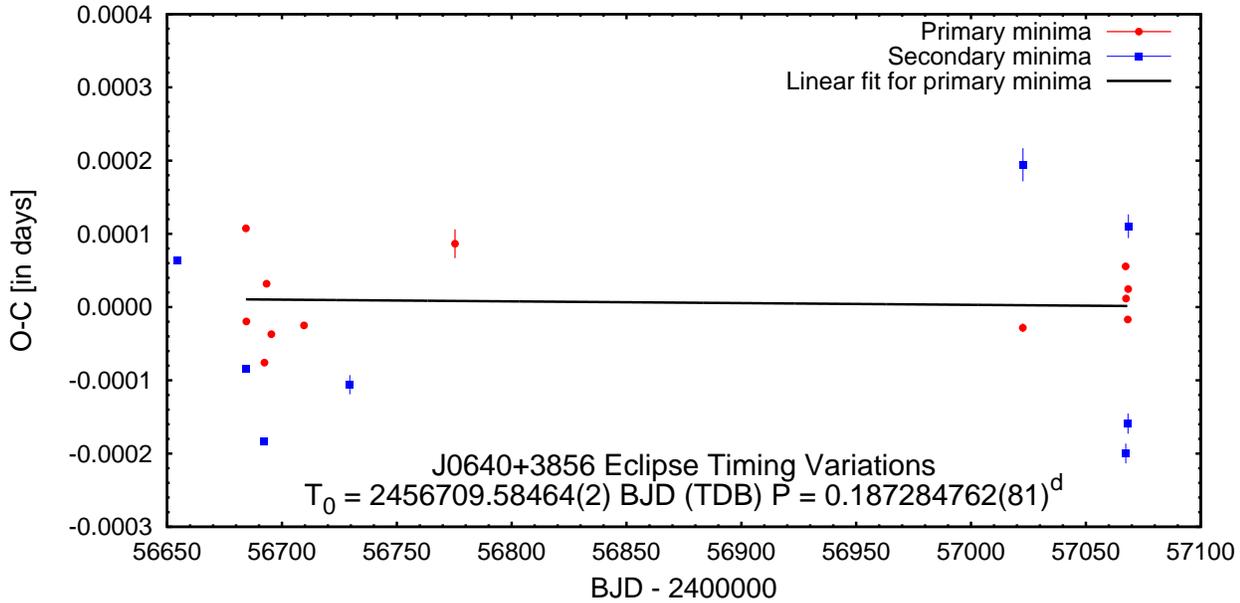}
\caption{ \label{Fig:ETV} Eclipse Timing Variation diagram of J0640+3856. Red circles represent the primary and blue boxes the secondary minima. For a better visibility, secondary minima with a large scatter are not shown here. The black line represents the linear fit for the primary minima, which we used to calculate the ephemeris and orbital period given both in the text and the figure. It shows that the period is well determined and constant over one year.}
\end{figure*}

We checked the stability of the orbital period. In order to determine
accurate times of minima the WHT light curves of the two primary and one
secondary minima were used as templates, and these template curves were fitted
to the other, less accurate light curves as described
in \citet{borkovitsetal15} in details. Thus, we obtained 13 primary and 14
secondary times of minima, which are listed in Table~\ref{Tab:ToM}, and plotted in Fig.~\ref{Fig:ETV}.  
For the period variation analysis, however, only the primary
minima were used, with the one exception of the outlying point at JD 2\,456\,691.
Therefore, we used twelve primary minima obtained between JD 2\,456\,684 and JD
2\,457\,068 to calculate the following ephemeris:
\begin{displaymath}
T_{\rm min,I}=2\,456\,709.58464(2)+0\fd187284762(81)\times E,
\end{displaymath}
where the epoch is given in BJD (TDB). We did not find any evidence for a
variation in the period, however, it should be noted that the scattering of the
individual points exceeds significantly the estimated statistical errors.

\section{Spectroscopy}

\subsection{Modelling the composite spectrum}{\label{Sec:spmodel}}

The first light curve of J0640+3856 clearly suggested a HW~Vir type eclipsing
binary with a strongly irradiated cool companion. In such binaries the orbit is
small and the companion is relatively large to produce a prominent reflection
effect in the light curve and provide full eclipses of the primary.

\begin{figure*}
\includegraphics[width=\textwidth]{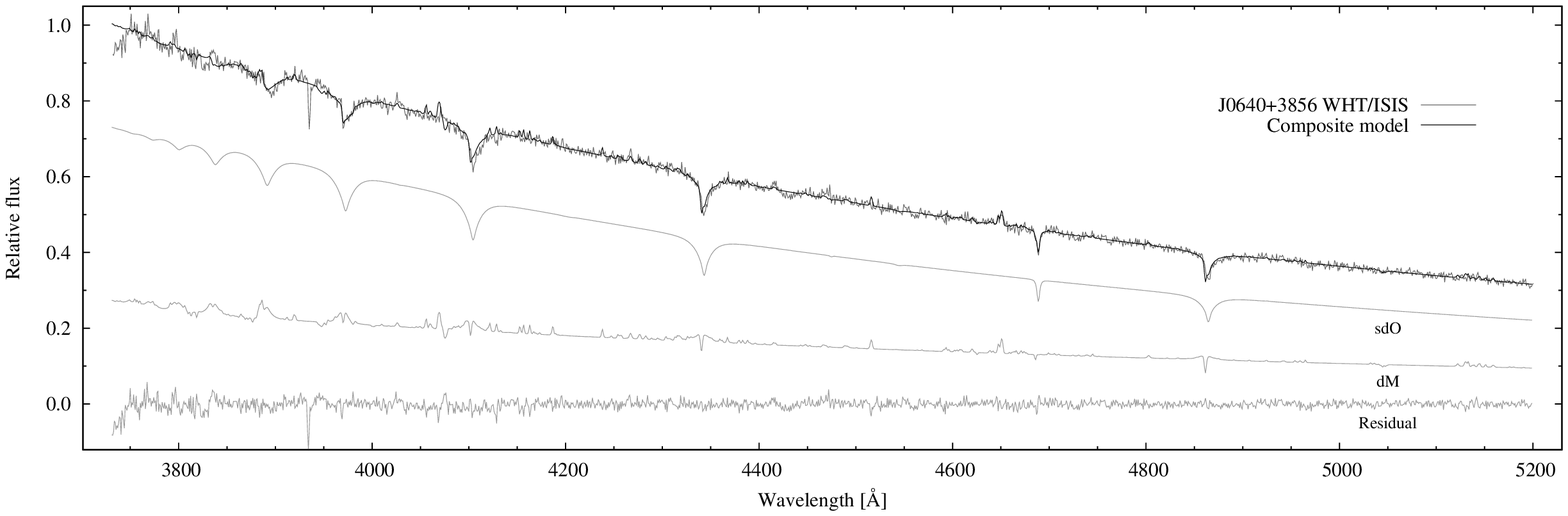}
\caption{ \label{fig:decomp} Spectral decomposition of the WHT/ISIS spectrum at
$\varphi=0.647$. The spectral features can be described with the superposition
of
a 55\,000\,K sdO star and an irradiated dM companion heated to
$\sim$22\,500\,K.}
\end{figure*}

Our classification spectra showed strong emission lines of C superimposed on the
shallow absorption spectrum of a hot star. J0640+3856 is therefore a double
lined spectroscopic binary, making it a target with high potential.
Disentangling the spectrum of the hot spot usually requires high-contrast
high-SNR spectroscopy and a challenging analysis procedure, similar to direct
exoplanet spectroscopy. The spectra of irradiated companions have been
disentangled and analysed in several hot subdwarf binaries, e.g.\ EC~11575 and 
V664~Cas \citep{exter05}, WD0137-349 \citep{maxted06} and AA Dor \citep{vuckovic08}. 
However, these
analyses were limited to relative line strength and radial velocity measurements
based on Gaussian line profiles. 
Although grids of irradiated M-dwarf spectra
are available (e.g.\ \citealt{barmanetal04}), their application to fit 
observed data is limited. 
\cite{wawrzyn09} presented a self-consistent model atmosphere analysis for the 
hot subdwarf binary UU Sge. 
Here we follow their methods based on the description in \cite{gunther11}.

J0640+3856 showed a double lined composite spectrum immediately in
low-resolution moderate-SNR spectroscopy, as achievable with 4\,m-class
telescopes, despite being fainter by $\sim$6 magnitudes in $V$ than AA~Dor. This
suggests that both components are more extreme: the sdO star must be hotter and
the companion must be a larger, probably earlier-M star that suffers a stronger
irradiation. The most remarkable spectral features are the distorted,
phase-dependent H Balmer line profiles, the strong He{\sc\,ii} absorption
indicating an sdO star and C{\sc\,ii-iii} lines in emission from the irradiated
companion. These features defined the starting models in our analysis. The WHT
observations include the secondary minimum, what makes them ideal to investigate
the irradiated hemisphere of the companion.

\begin{figure*}
\includegraphics[width=\textwidth]{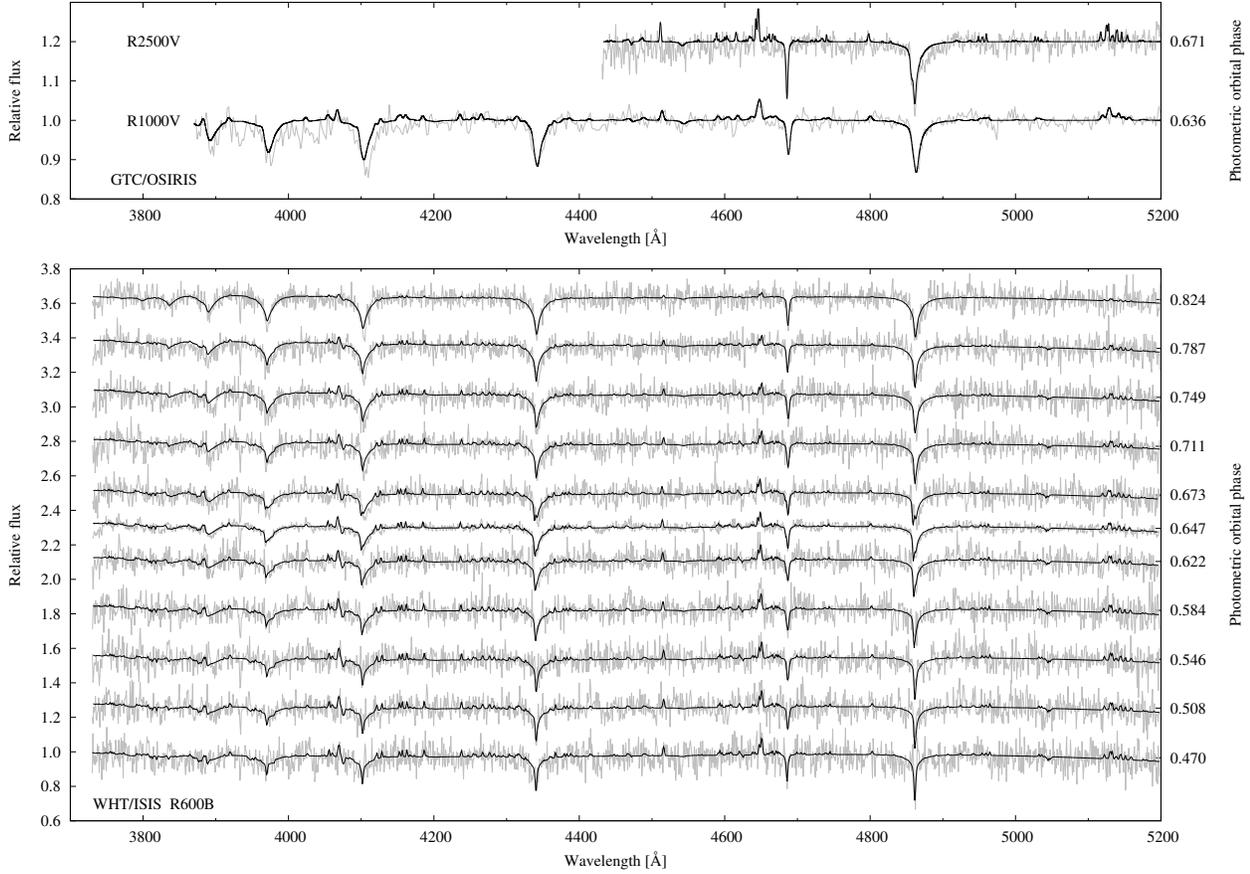}
\caption{ \label{fig:wht}{\sl Top:} GTC/OSIRIS spectra of J0640+3856 (grey) with
best fit models (black).
{\sl Bottom:} Best fit models (black) for the phase-dependent radial velocity
corrected composite spectra
obtained with WHT/ISIS (grey). We applied a spectral decomposition of the middle
spectrum at orbital phase
$\varphi=0.647$.  All the other models were calculated from this decomposition
by shifting the components
in radial velocity and scaling them by the orbital phase to account for the
visibility of the substellar
point of the companion.}
\end{figure*}

Our spectral analysis was based on the steepest-descent iterative binary fit
procedure, {\sc XTgrid}, developed by \cite{nemeth12}. {\sc XTgrid} employs {\sc
Tlusty/Synspec} \citep{hubeny95,lanz03} non-LTE model atmospheres and synthetic
spectra to reproduce composite binary spectra. The strong irradiation allowed us
to model both components in J0640+3856 with {\sc Tlusty}. The model atmospheres
include H, He, C, N and O opacities consistently in the atmosphere structure and
synthetic spectrum calculations for both stars. Although we could not identify
CNO absorption lines in the sdO star and kept their abundances fixed at
$\log(n{\rm X}/n{\rm H})=-6$, we included CNO opacities, because they have an
effect on the temperature structure of non-LTE atmosphere models
\citep{werner96}.

We started an {\sc XTgrid} process for the sdO star and another for an M dwarf
that is irradiated by the sdO. These processes iteratively updated the stellar
parameters and fit the observed composite spectrum together. As a first
approximation we applied an isotropically irradiated model for the inner
hemisphere of the companion. We assumed a black-body energy distribution of the
irradiating flux with appropriate temperature for the sdO. The geometric
dilution factor ($W$) is \citep{mihalas78}:
\begin{displaymath}
W = \frac{1}{2} \left( 1-\sqrt{1- \left( {\frac{{\rm R}_2}{a}} \right)^2 }
\right),
\end{displaymath}
where ${\rm R}_2$ is the radius of the secondary and $a$ is the semi-major axis.
The dilution factor describes the strength of irradiation. We used an empirical
dilution factor $W_e=0.05$ that was found for the substellar point in the
similar system AA~Dor. The empirical dilution factor is a fudge factor in our
analysis and substantially larger than the geometrical dilution factor. The
difference is probably due to the non-planckian sdO spectrum and due to the fact
that the spectrum of the irradiated companion cannot be fully described with
the conditions in the substellar point, as suggested by \citet{gunther11}. 

The combined spectrum at $\varphi=0.647$ has a SNR of $\sim45$, therefore we
modelled this spectrum as shown in Fig.~\ref{fig:decomp}. The flux contribution
of the hot spot was determined for this spectrum and we approximated it for the
other spectra by scaling with orbital phase according to $F_{\rm dM}/F_{\rm
sd}\sim0.43\,\sin^2 \left({\pi\,\varphi}\right)$. The {\sc Tlusty} model
provides the temperature structure of the irradiated companion. The temperature
decreases steeply inward, reaches a minimum and progressively increases again.
The minimum temperature is $T=22\,500$\,K at optical depth $\tau\approx0.6$,
which we associate with the photospheric effective temperature of the substellar
point. The parameters derived from the spectroscopic modelling are listed in
Table \ref{Tab:syntheticfit} and suggest an sdO+dM binary. 

Next we shifted the components in velocity space to reproduce the
phase-dependent composite spectra. Fig.~\ref{fig:wht} shows the best fits for
the ten WHT blue arm observations. These fits show that all H  line profiles are
contaminated by the emission lines from the companion and the relative strength
of this contamination increases towards the Balmer series limit. 
The inverted temperature structure of the companion results in high excitation 
lines compared to the photospheric temperature. Such emission lines of 
C\,{\sc\,iii} and Si\,{\sc\,iii} have been observed from the irradiated companion 
in other systems (e.g.: V477 Lyr, AA Dor). Therefore, a He\,{\sc\,ii} $4686$\,\AA\ 
emission line may be expected as well. 
However, this line forms between very high lying levels in the He ion and at the 
low He abundance of J0640+3856 we did not find any contribution. 
The fact that the He{\sc\,ii}
$4686$\,\AA\ line forms exclusively in the sdO star allowed us to measure its 
radial velocity.
The C{\sc\,ii}
$4267$\,\AA\ line, the C{\sc\,iii} blend near $4650$\,\AA\ and $4070$\,\AA\ are
in emission and come from the irradiated companion.  Based on a selection of
these lines we could also estimate the radial velocity of the companion.

\subsection{Radial velocity and stellar masses}
\label{rv}

Even though precision radial velocity measurements would require high-dispersion
and higher-SNR spectra, we attempted to estimate radial velocities from the
WHT/ISIS spectra, because the amplitudes put a useful constraint on the mass
ratio. First, we checked the dispersion correction, which was based on 150 lines
of the CuAr and CuNe calibration lamps, and tested the stability on the
Ca{\sc\,ii}\,K line. This line forms mostly in the interstellar medium and
therefore its radial velocity can be used to check dispersion offsets. We found
the K line to be consistent in all our spectra, but systematically offset by
40\,km\,s$^{-1}$. Most of this offset can be explained with the barycentric
velocity correction $v_{BC}=-26.5$ km\,s$^{-1}$ during the observations. The
interstellar extinction towards J0640+3856 is $E({\rm B-V})=0.124\pm0.005$
\citep{schlafly11}, therefore we associate the remaining radial velocity shift
of the Ca{\sc\,ii}\,K line with the relative velocity of the interstellar
material with respect to the Solar system. We performed two independent
experiments to measure velocities and limited our measurements to the
He{\sc\,ii}  $4686$\,\AA\ line for the sdO and to the C{\sc\,iii} blend
between $4647$--$4652$\,\AA\ for the dM companion as these are the strongest
undistorted features in the spectra.

In the first method we combined each consecutive spectrum, like those at
$\varphi=0.622$ and $0.673$ to get the one at $0.647$ in Fig.~\ref{fig:wht}, and
applied a Savitzky-Golay filter \citep{sav64}. Then we measured the SNR of each
of these combined spectra. Next, we determined the radial velocity with an
iterative chi-square minimization cross-correlation procedure.  We decided to
use the preceding spectrum as a self-template to reduce systematic effects in
the chi-square method. For each observed spectrum we performed 100 radial
velocity measurements to obtain a mean value and standard deviation. We
resampled the spectra according to the SNR before each individual measurement.

The second method was based on visual inspection. This allowed us to compare the
observations to synthetic spectra directly and derive absolute radial
velocities.

We found a good match between the two methods for the sdO star. However, for the
dM companion we found that the cross-correlation method underestimated the
radial velocity by about 40\%. The most plausible reason for this is orbital
smearing that makes the spectral lines considerably shallower and broader, while
the decreasing contribution of the companion also makes the last data points
less reliable. Similarly for the sdO star, the last data points show a
decreasing radial velocity in conflict with the photometric orbital period. We
attributed these inconsistencies to a lower data quality and disregarded the
last three spectra in the radial velocity measurement. The symmetric appearance
of the reflection effect and secondary minimum suggest a circular orbit and a
co-rotating companion. Assuming synchronised rotation we found a projected
rotational velocity $v_{\rm rot}\sin{i}=49.2$\,km\,s$^{-1}$ using $R_2=0.2$
R$_\odot$ from the light curve solution (see Sect.~\ref{phot}). 
As the centre of mass of the companion
does not coincide with its centre of light we needed to correct the radial
velocity with the projected rotational velocity at the centre of light. We
calculated this correction as $-29$\,km\,s$^{-1}$ at $\varphi=0.75$.

\begin{figure}[h]
\includegraphics[width=\linewidth]{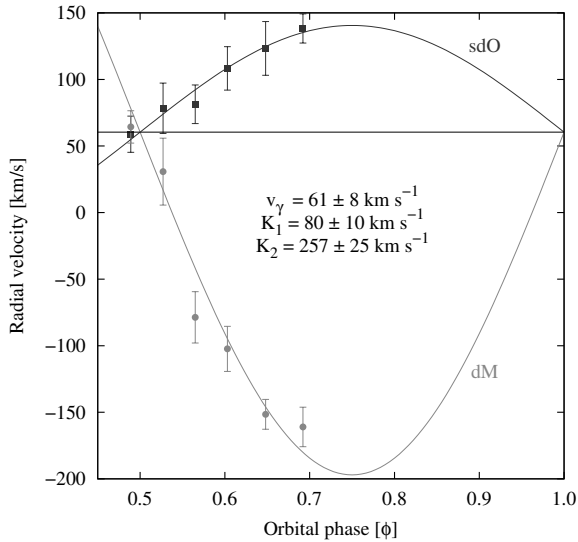}
\caption{\label{fig:rvc}Radial velocity curves of the components of J0640+3856.}
\end{figure}

The six data points in Fig.~\ref{fig:rvc} show a clear trend and suggest the
corrected radial velocity semi-amplitudes: $K_1=80\pm10$\,km\,s$^{-1}$,
$K_2=257\pm25$\,km\,s$^{-1}$ and a barycentric system velocity
$v_{\gamma}=61\pm8$\,km\,s$^{-1}$.
From the radial velocity semi-amplitudes we found a mass ratio of
$q=M_2/M_1=K_1/K_2=0.31\pm0.04$. Assuming a circular orbit, the radial velocity
semi-amplitudes define the projected semi-major axis
$a\sin{i}\approx\left(K_1+K_2\right)\,P/2\pi$, where $P$ is the orbital period
and $i$ is the inclination, which can be obtained from the eclipsing light curve
solution. The observed orbital period and this semi-major axis define the total
mass by Kepler's third law. 

Using the inclination from the light curve solution $i=87.11^{\circ}$ 
and the radial velocity amplitudes the semi-major axis is $a= 0.006$\ AU ($869\,000$ km).
Then the total mass from Kepler's third law is ${\rm M}_{tot}=0.744$ ${\rm M}_\odot$.

According to our models the strong irradiation transforms the secondary atmosphere completely. The irradiated models are insensitive for the unperturbed temperature of the companion, therefore we cannot assess the night side temperature from the currently available irradiated spectrum. In turn, the models show that the Balmer emission lines are sensitive for the surface gravity (pressure) and to reproduce the line profile variations they require a $\log{g}>4.5$. Assuming that the mass of the subdwarf is close to the canonical 0.5 M$_\odot$, the measured surface gravity $\log{g}=4.9$ of the companion is most consistent with a mid-M type star.

\section{Light curve analysis}
\label{phot}

We carried out a three-band simultaneous light curve analysis with the recently
developed {\sc Lightcurvefactory} code \citep{borkovitsetal13,borkovitsetal14}.
We chose the Sloan $i'$ and $r'$ light curves obtained using WHT/ACAM, which
contain two flat primary minima offering strong geometrical constraints not only
upon the inclination and the relative radii of the stars, but also, via Kepler's third law,
the surface gravity of the stars given an approximate estimate of their mass.
The third light curve used was the $R_C$-band dataset obtained with the Schmidt
telescope at the Piszk\'estet\H o Observatory, which covers more than a full
orbit of the system, including two primary minima. Despite a lower quality and
the absence of measurements during the deepest parts of the primary minima,
these data significantly improved the fit of the reflection effect and also
contain useful information on the ellipsoidal variation or its absence that can be used to
further constrain the mass ratio.

\begin{figure*}
\includegraphics[width=\hsize]{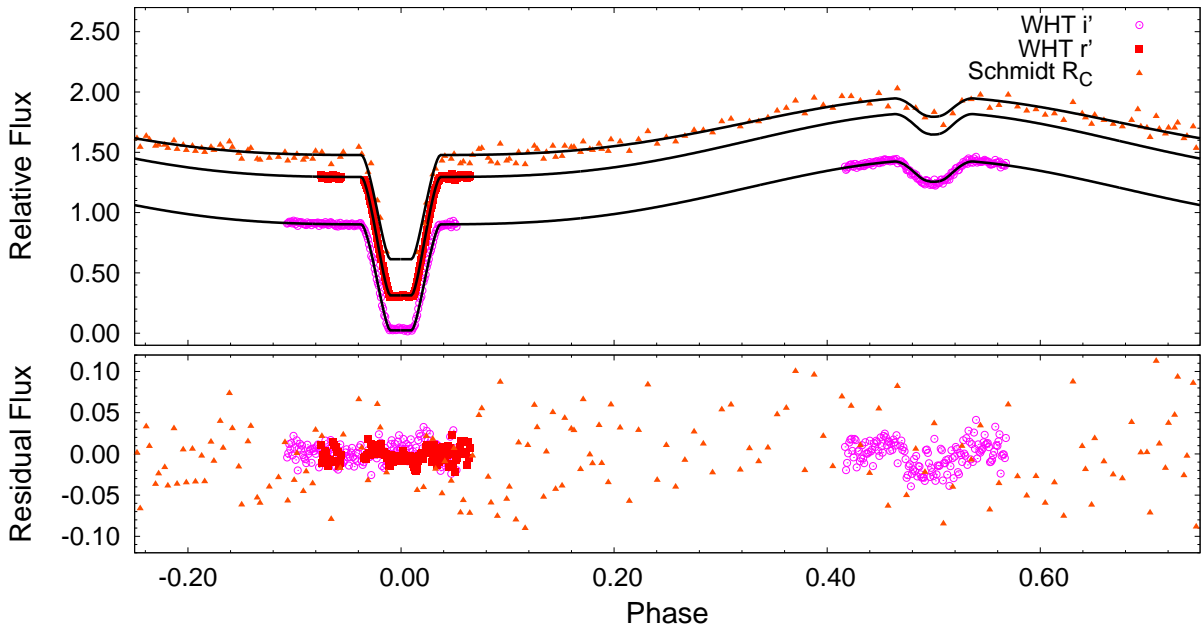}
\caption{\label{lcfit} Observed light curves and solutions with their residuals
for the
Sloan $i'$, $r'$ (WHT) and $R_{C}$ (Piszk\'estet\H o Observatory) passband
measurements.}
\label{fig:lcfit}
\end{figure*}

There is no indication for eccentricity from the spectroscopic orbit or the
phase of secondary eclipse so we assumed that the orbit is circular, as expected
for such a short-period binary star. Assuming a spherical primary, and a marginally oblated secondary
star (which latter assumption was found to be reliable from a preliminary analysis of the eclipse geometry),
it is a suitable approximation that
the fractional radii of the components relate directly to the observable
quantities of the full and totality-phase durations of the primary occultations.
In such a way the whole system geometry was determined except for one free
parameter, namely $i$. Therefore we were able to express the adjustable
dynamical and geometrical parameters using the binary period ($P$), epoch
($T_0$), the orbital inclination ($i$) and the mass ratio ($q$). Although the
mass ratio is also known from the radial velocity solution (see Sect.~\ref{rv}),
we made fits both with it freely adjusted and fixed to the spectroscopic value,
as a consistency check.  Due to the detached configuration and nearly spherical
stellar shapes, the mass ratio has only a minor influence on our light curve
solution.

Considering the atmospheric properties, the effective temperature of the sdO
primary was fixed to the value obtained from the spectroscopic analysis
($T_\mathrm{eff1}=55\,000$K), while $T_\mathrm{eff2,d}$ (night side of the
secondary component) was the fifth adjusted parameter. The other atmospheric
parameters -- limb darkening, gravity brightening coefficients, bolometric
albedos and abundances of the sdO primary -- were also kept fixed. For limb
darkening we applied the logarithmic law, and the coefficients were calculated
according to the passband-dependent precomputed tables\footnote{Downloaded from
the site http://phoebe-project.org/1.0/} of the {\sc phoebe} team
\citep{prsazwitter05,prsa11} which were based on the tables of
\citet{castellikurucz04}.

On the other hand, most of the atmospheric parameters of the secondary star were
involved into the fitting process. This was done because of the high irradiation
which results in such a high temperature as $T=22\,500$~K at the substellar
point that there are significant deviations from LTE models as discussed e.g.\
in \citet{barmanetal04}. Amongst other consequences, this may change the full
atmosphere to a radiative one, as was observed for other similar systems (e.g.\
AA~Dor; \citealt{hilditchetal03}), and may even result in limb-brightening
(i.e.\ negative limb darkening coefficients, see e.g.\ \cite{pol94a}, for
V477~Lyr).

Finally, the passband luminosities of the primary component were also
re-calculated in each step. A $\chi^2$ minimization was carried out using
firstly the Levenberg-Marquardt differential corrections algorithm, and then a
final refinement with a grid-search method. The resulting light curve solutions, and their residual curves are shown in Fig.~\ref{fig:lcfit}. The values of the free parameters
found from the best fit to the light curves are given in
Table~\ref{Tab:syntheticfit}, together with the fixed parameters in the
least-squares fit and some useful derived parameters.

\begin{table*}
\begin{minipage}{\textwidth}
\begin{center}
\centering
\caption{Stellar and orbital parameters derived from the spectroscopic and
eclipsing light curve analysis.}
 \label{Tab:syntheticfit}
 \begin{tabular}{@{}lll}
  \hline
& \multicolumn{2}{c}{orbital parameters} \\
\hline
  $P_\mathrm{orb}$ (days)  & \multicolumn{2}{c}{$0.18728550\pm0.00000005$} \\
  $T_\mathrm{MIN I}$ (BJD) & \multicolumn{2}{c}{$2\,456\,709.584565\pm0.000013$}
\\
  $a^d$ (R$_\odot$)        & \multicolumn{2}{c}{$1.24857\pm0.09976$} \\
  $e$                      & \multicolumn{2}{c}{$0.0$}  \\
  $i$ ($\degr$)            & \multicolumn{2}{c}{$87.11\pm0.03$}  \\
  $q_\mathrm{spec}^d$      & \multicolumn{2}{c}{$0.31\pm0.05$} \\
  $q_\mathrm{phot}$ (unused)& \multicolumn{2}{c}{$0.19\pm0.02$} \\
%  $K$ (kms$^{-1}$)         & $80\pm10$ & $228\pm26$ \\
%  $V_\gamma$ (kms$^{-1}$) & \multicolumn{2}{c}{$61\pm8$} \\
\hline
&\multicolumn{2}{c}{stellar parameters} \\
\hline
   & {\bf Primary} & {\bf Secondary} \\
  \hline
&\multicolumn{2}{c}{fractional radii$^d$} \\
  \hline
 $r_\mathrm{pole}$  & $0.07646\pm0.00053$ & $0.15703\pm0.00055$ \\
 $r_\mathrm{side}$  & $0.07646\pm0.00053$ & $0.15838\pm0.00055$ \\
 $r_\mathrm{point}$ & $0.07646\pm0.00053$ & $0.16203\pm0.00065$ \\
 $r_\mathrm{back}$  & $0.07646\pm0.00053$ & $0.16130\pm0.00063$ \\
 \hline
&\multicolumn{2}{c}{absolute stellar parameters} \\
  \hline
 $M^d$ (M$_\odot$)            & $0.567\pm0.138$                 &
$0.177\pm0.051$  \\
 $R^d$ (R$_\odot$)            & $0.0955\pm0.0077$               &
$0.1985\pm0.0159$ \\
 $T_\mathrm{eff}$ (spec) (K)  &$55\,000\pm3\,000$               &
$4000^{+1000}_{-1500}$ \\
 $T_\mathrm{eff}$ (phot) (K)  &$55\,000$ $(\pm3\,000)$          & $4648\pm55$
$(\pm259)$ \\
 $L_\mathrm{bol}^d$(L$_\odot$)& $73.692\pm11.819$ $(\pm19.955)$ &
$0.016\pm0.004$ $(\pm0.007)$ \\
 $\log g$ (spec) (cgs)        & $5.97\pm0.30$                   & $4.9\pm0.5$ \\
 $\log g^d$ (phot) (cgs)      & $6.23\pm0.04$                   & $5.11\pm0.07$
\\
 \hline
&\multicolumn{2}{c}{chemical abundances from spectroscopy} \\
  \hline
$\log (n\mathrm{He}/n\mathrm{H})$\   &$-2.24\pm0.4$   & $-2.4>$    \\
$\log (n\mathrm{C}/n\mathrm{H}) $\   &$-6.0$          & $-2.0>$    \\
$\log (n\mathrm{N}/n\mathrm{H}) $\   &$-6.0$          & $-5.0>$    \\
$\log (n\mathrm{O}/n\mathrm{H}) $\   &$-6.0$          & $-3.5>$    \\
  \hline
&\multicolumn{2}{c}{flux ratio from spectroscopy} \\
  \hline
$(\rm{F}_{\rm dM}/\rm{F}_{\rm sd})_{\rm max}$ at 4500\,\AA
&\multicolumn{2}{c}{ $0.43\pm0.06$}\\
 \hline
&\multicolumn{2}{c}{atmospheric model-dependent parameters} \\
  \hline
$x_\mathrm{bol}$     & $0.231$ & $0.300\pm0.01$  \\
$y_\mathrm{bol}$     & $0.148$ & $-$             \\
$A$                  & $1.0$   & $1.09\pm0.02$   \\
$\beta$              & $1.0$   & $2.80\pm0.70$   \\
$x_{i}$              & $0.160$ & $0.47\pm0.02$   \\
$y_{i}$              & $0.108$ & $-$             \\
$L_i/(L_1+L_2) (i)$  & $0.827$ & $0.173$         \\
$x_{r}$              & $0.184$ & $0.50$          \\
$y_{r}$              & $0.122$ & $-$             \\
$L_i/(L_1+L_2) (r)$  & $0.986$ & $0.014$         \\
$x_{R_C}$            & $0.179$ & $0.50\pm0.02$   \\
$y_{R_C}$            & $0.119$ & $-$             \\
$L_i/(L_1+L_2) (R_C)$& $0.984$ & $0.016$         \\
\hline
\end{tabular}
\end{center}
\vspace{-6pt}
{\small
{\bf Notes.} (1) Parameters without uncertainties were kept fixed, or adopted
from precomputed tables. (2) Parameters subscripted with $d$ are derived
parameters. (3) Second uncertainties in parentheses were calculated setting the
uncertainty of the (fixed) primary effective temperature to be $\delta
T_\mathrm{eff1}=3\,000$K (i.e.\ its spectrocopic uncertainty). (4) $x$, $y$, $A$
and $\beta$ denote linear and logarithmic limb darkening coefficients,
bolometric albedos and gravity brightening exponents, respectively. (5) In the
case of the passband-dependent fractional luminosities $[L_i/(L_1+L_2)]$, the
reflection/irradiation effect was taken into account.}
\end{minipage}
\end{table*}

The photometric mass ratio was found to be significantly smaller
($q=0.19\pm0.02$) than the spectroscopic one. On the other hand, comparing the
two solutions (with adjusted or fixed mass ratios), we found that all the other
adjusted parameters, and also the $\chi^2$ have remained within their 1$\sigma$
values. This is a consequence of the  small amplitude ellipsoidal variation, due to
the minor oblateness of the secondary star and, therefore, the weak dependence of the light curve on $q$.
We therefore decided to keep the solution obtained with the fixed spectroscopic
mass ratio and the derived astrophysical quantities were computed accordingly.
Considering the other quantities common to the spectroscopic and photometric
analyses, while the light curve solution clearly confirms the spectroscopic
temperature of the secondary, the local gravities were found to be higher.

Using the spectroscopic mass ratio we found a mass of $0.57 \pm 0.14$ $M_{\odot}$ for the sdO
star and $0.18 \pm 0.05$ $M_{\odot}$ for the companion, in agreement with a
mid-M dwarf (M6V) classification. We note that although the spectroscopic mass
ratio is higher, yet consistent with our light curve analysis, the poor quality
of our spectra means it is not precisely determined.

Comparing the spectroscopic and photometric $\log g$ values of the sdO primary,
despite that the lower spectroscopic value is closer to expectations, the
photometric result is evidently the more robust. This is because it is
determined purely by the system geometry via the eclipse durations, which give
the relative radii as a the function of $i$. Then, by the use of Kepler's third
law, one can see that $g_\mathrm{pri}\sim m_\mathrm{pri}^{1/3}/(1+q)^{2/3}$.
Therefore, even a $100\%$ error in the mass of the primary would result in a
$0.1$ dex discrepancy in its $\log g$ value. This is valid only for spherical
stars with negligible tidal and rotational effects, but our solution (i.e.\ the
low values of the fractional radii)  is consistent with the assumption that
these effects play only a minor role in the system. As a consequence, we accept
the high $\log g$ values obtained from the light curve analysis instead of their
spectroscopic value and conclude that the sdO primary is a compact object on its
way to the WD cooling sequence.

Turning to the adjusted atmospheric parameters of the secondary, its bolometric
albedo ($A_2$) was found to be greater than unity. This is not an unphysical
solution, but implies that light from outside of a given photometric passband is
being reprocessed and re-emitted in the passband. A similar situation has been
found before for other systems (\citealp[e.g.,][for KIC~10661783, a totally
eclipsing binary with a $\delta$ Scuti component]{southworthetal11} and
\citealp[][for the eclipsing sdOB+dM binary V1828~Aql]{almeidaetal12}). We also
found an unusually high gravity darkening coefficient ($\beta_2$) for the
secondary. The large uncertainty, however, makes this result ambigous. 
 We note that the relatively poor fit to the data in secondary minimum is likely due to the simplistic treatment of the most highly-irradiated part of the atmosphere of the secondary component, which is eclipsed during secondary minimum. A complete physical description of the phsyics in this atmosphere is not available in our code, or in other commonly available light curve modelling codes.

\section{Summary and Conclusions}

We have discovered an sdO+M6V eclipsing binary ({\sc RA}=$\rm{06^{h}40^{m}29^{s}11}$; {\sc
 Dec}=$\rm{+38^{\circ}56\arcmin52\arcsec2}$; J=2000.0; R$_{\rm
{max}}$=17.2 mag) that shows 6-mag deep primary
eclipses and a $\sim$0.5\,mag reflection effect. These are the most extreme
variations among all HW~Vir type binaries known to us and the primary minimum is
even deeper than that of NN~Ser, a well-known white dwarf with an extreme
eclipse depth \citep{hae04}. With photometric and spectroscopic follow-up we
constrained the atmospheric properties of the components and the binary orbit.
Although the specific spectral features and the effective temperature
($T_\mathrm{eff1}=55\,000$ K) classify the primary component as an sdO star, the
surface gravity is at the upper limit of sdOs ($\log g = 6.2$ cgs) and the
radius  ($R=0.096\,R_{\odot}$) is smaller than for normal sdO stars. These
parameters place the primary component in a special position, suggesting the sdO
star is a pre-WD, similar to BE~UMa which is classified as a borderline object
between sdO subdwarfs and DAO white dwarfs \citep{fer99}. The non-detection of a
PN around J0640+3856 also supports the evolved hot subdwarf (post-sdB)
scenario. 

We have constructed a simple model to reproduce the spectral contribution of 
the irradiated companion. Although this model is optimized for the substellar 
point, it represents the day side of the companion well, suggesting that the 
strong irradiation heats up the entire inner hemisphere homogeneously.

Our results suggests that the secondary component may be inflated by only a few
percent, like in the cases of similar close binaries \citep{afs08}. We estimate
that the substellar point of the red dwarf is heated to about 22\,500\,K. The
heat transport of these inflated stars is ineffective, so the large temperature
difference between the day and night sides is preserved over long timescales \citep{rit00}.

We conclude that the most probable companion spectral type is mid-M. A later-type or more compact companion would be unable to reproduce the eclipses while an earlier type and more massive companion would be inconsistent with the radial velocity curve.

The biggest advantage of J0640+3856 is that it is a double-lined spectroscopic
binary. Spectroscopic observations covering the full orbital cycle will yield
more precise parameters (especially masses) for the components, as well as an
opportunity to monitor and analyse the changing features in the spectra caused
by the reflection effect.

The J0640+3856 system is a good analogue to study interactions in planetary
systems with hot Jupiters. Both the illumination effect in the primary
minimum
and the thermal radiation and reflected light disappearance and reappearance in
the secondary minimum, is similar as is the luminosity ratio.

The newly discovered J0640+3856 is a unique laboratory in several aspects and
opens opportunities to make further steps to understand the evolutionary history
of post-common envelope binaries.

\acknowledgments
This project has been supported by the Hungarian OTKA Grants K83790, K104607,
K109276, K113117, ESA PECS Contract No. \linebreak 4000110889/14/NL/NDe, the
Lend\"ulet-2009 and the Lend\"ulet LP2012-31 Young Researchers Programme of the Hungarian Academy of Sciences
and the European Community's Seventh Framework Programme (FP7/2007-2013) under
grant agreement no. 269194 (IRSES/ASK) and no. 312844 (SPACEINN). AD has been
supported by the Postdoctoral Fellowship Programme of the Hungarian Academy of
Sciences and the J\'anos Bolyai Research Scholarship of the Hungarian Academy of
Sciences. PN was supported by the Deutsche Forschungsgemeinschaft under grant He
1356/49-2. JS acknowledges financial support from STFC in the form of an
Advanced Fellowship. TB would like to thank City of Szombathely for support
under Agreement No. S-11-1027. Based on observations made with the Gran
Telescopio Canarias (GTC), instaled in the Spanish Observatorio del Roque de los
Muchachos of the Instituto de Astrofísica de Canarias, in the island of La
Palma.

{\it Facilities:} \facility{WHT, GTC}.


\begin{thebibliography}{}


\expandafter\ifx\csname natexlab\endcsname\relax\def\natexlab#1{#1}\fi

\bibitem[Af\c{s}ar \& \.{I}bano\u{g}lu(2008)]{afs08} Af\c{s}ar, M., \&
\.{I}bano\u{g}lu, C., 2008, \mnras, 391, 802

\bibitem[Aller et al.(2015)]{all15} Aller, A., Miranda, L.~F., Olgu{\'{\i}}n,
L., et al.\ 2015, \mnras, 446, 317 

\bibitem[Almeida et al.(2012)]{almeidaetal12} Almeida, L. A., Jablonski, F.,
Tello, J., \& Rodrigues, C. V., 2012, \mnras, 423, 478

\bibitem[Barman et al.(2004)]{barmanetal04} Barman, T. S., Hauschildt, Peter H.,
\& Allard, F., 2004, \apj, 614, 338

\bibitem[Benn, Dee \& Ag\'ocs(2008)]{ben08}
 Benn, C., Dee, K., Ag\'ocs, T., 2008, Ground-based and Airborne Instrumentation for Astronomy II. Edited by McLean, Ian S.; Casali, Mark M. Proceedings of the SPIE, 7014, 6

\bibitem[Borkovits et al.(2013)]{borkovitsetal13} Borkovits, T., Derekas, A.,
Kiss, L. L., et al., 2013, \mnras, 428, 1656

\bibitem[Borkovits et al.(2014)]{borkovitsetal14} Borkovits, T., Derekas, A.,
Fuller, J., et al., 2014, \mnras, 443, 3068

\bibitem[Borkovits et al.(2015)]{borkovitsetal15} Borkovits, T., Rappaport, S., 
Hajdu, T., Sztakovics, J., 2015, \mnras, 448, 946

\bibitem[Castelli \& Kurucz(2004)]{castellikurucz04} Castelli, F., \& Kurucz, R.
L., 2004, [arXiv:astro-ph/0405087]

\bibitem[Exter et al.(2005)]{exter05} Exter, K.~M., Pollacco, D.~L., Maxted,
P.~F.~L., Napiwotzki, R., \&
Bell, S.~A.\ 2005, \mnras, 359, 315

\bibitem[Ferguson et al.(1999)]{fer99}
 Ferguson, D. H., Liebert, J., Haas, S., Napiwotzki, R., James, T. A., 1999,
\apj, 518, 866

\bibitem[G{\"u}nther \& Wawrzyn(2011)]{gunther11} G{\"u}nther, H.~M., \&
Wawrzyn, A.~C.\ 2011, \aap,
526, AA117

\bibitem[Haefner et al.(2004)]{hae04}
 Haefner, R.; Fiedler, A.; Butler, K.; Barwig, H., 2004, \aap, 428, 181

\bibitem[Han et al.(2002)]{han02} Han, Z., Podsiadlowski, P., Maxted, P.~F.~L.,
Marsh, T.~R., \& Ivanova, N., 2002, \mnras, 336, 449

\bibitem[Han et al.(2003)]{han03} Han, Z., Podsiadlowski, P., Maxted, P.~F.~L.,
\& Marsh, T.~R., 2003, \mnras, 341, 669

\bibitem[Heber(2009)]{heb09} Heber, U., 2009, ARA\&A, 47, 211

\bibitem[Hilditch et al.(2003)]{hilditchetal03} Hilditch, R. W., Kilkenny, D.,
Lynas-Gray, A. E., \& Hill, G., 2003, \mnras, 344, 644

\bibitem[Hubeny \& Lanz(1995)]{hubeny95} Hubeny, I., \& Lanz, T., 1995, \apj,
439, 875

\bibitem[Kawka et al.(2015)]{kaw15} Kawka A., Vennes S., O'Toole S. et al.,
2015, MNRAS,  in press

\bibitem[Lanz \& Hubeny(2003)]{lanz03} Lanz, T., \& Hubeny, I., 2003, \apjs,
146, 417

\bibitem[Maxted et al.(2001)]{max01} Maxted, P. F. L., Heber, U., Marsh, T. R.,
\& North, R. C., 2001, MNRAS, 326,1391

\bibitem[Maxted et al.(2006)]{maxted06} Maxted, P.~F.~L., Napiwotzki, R.,
Dobbie, P.~D., \& Burleigh, M.~R.\ 2006, \nat, 442, 543

\bibitem[Mihalas (1978)]{mihalas78} Mihalas, D. 1978, Stellar Atmospheres, 2nd
edn. (San Francisco:
Freeman)

\bibitem[Napiwotzki et al.(2004)]{nap04} Napiwotzki, R., Karl, C. A., Lisker,
T., Heber, U., Christlieb, N., Reimers, D., Nelemans, G., \& Homeier, D., 2004,
Ap\&SS, 291, 321

\bibitem[N{\'e}meth et al.(2012)]{nemeth12} N{\'e}meth, P., Kawka, A., \&
Vennes, S.\ 2012, \mnras, 427, 2180

\bibitem[Paczynski(1976)]{pac76} Paczynski, B., 1976, Structure and Evolution of
Close Binary Systems, 73, 75

\bibitem[Pollacco \& Bell(1994)]{pol94a} Pollacco, D. L., \& Bell, S. A., 1994,
\mnras, 267, 452

\bibitem[Pollacco, Bell \& Hilditch(1994)]{pol94b} Pollacco, D. L., Bell, S. A.,
\& Hilditch, R. W., 1994, \mnras, 270, 449

\bibitem[Pr{\v s}a(2011)]{prsa11} Pr{\v s}a, A., 2011, PHOEBE Scientific
Reference, Dept. of Astronomy and Astrophysics, Villanova University

\bibitem[Pr{\v s}a \& Zwitter(2005)]{prsazwitter05} Pr{\v s}a, A., \& Zwitter,
T., 2005, \apj, 628, 426

\bibitem[Ritter, Zhang \& Kolb(2000)]{rit00}
 Ritter, H., Zhang Z.-Y., Kolb U., 2000, \aap, 360, 969

\bibitem[Savitzky \& Golay(1964)]{sav64} Savitzky, A., \& Golay, M.~J.~E.\ 1964,
Analytical Chemistry, 36, 1627

\bibitem[Schlafly \& Finkbeiner(2011)]{schlafly11} Schlafly, E.~F., \&
Finkbeiner,
D.~P.\ 2011, \apj, 737, 103

\bibitem[Southworth et al.(2011)]{southworthetal11} Southworth, J., Zima, W.,
Aerts, C. et al., 2011, \mnras, 414, 2413

\bibitem[Stroeer et al.(2007)]{str07} Stroeer, A., Heber, U., Lisker, T.,
Napiwotzki, R., Dreizler, S., Christlieb, N., \& Reimers, D., 2007, A\&A, 462,
269

\bibitem[Taam \& Ricker(2006)]{taam06} Taam, R.~E., \& Ricker, P.~M.\ 2006, arXiv:astro-ph/0611043 

\bibitem[Vu{\v c}kovi{\'c} et al.(2008)]{vuckovic08} Vu{\v c}kovi{\'c}, M.,
{\O}stensen, R., Bloemen, S., Decoster, I.,
\& Aerts, C., 2008, Hot Subdwarf Stars and Related Objects, 392, 199

\bibitem[Wawrzyn et al.(2009)]{wawrzyn09} Wawrzyn, A.~C., Barman, T.~S., G{\"u}nther, H.~M., Hauschildt, P.~H., \& Exter, K.~M.\ 2009, \aap, 505, 227 

\bibitem[Werner(1996)]{werner96} Werner, K.\ 1996, \apjl, 457, L39


\end{thebibliography}
\end{document}